\PassOptionsToPackage{table,xcdraw}{xcolor} 
\documentclass[sigconf,screen]{acmart}

\acmConference[ICSE 2022]{The 44th International Conference on Software Engineering}{May 21–29, 2022}{Pittsburgh, PA, USA}

\usepackage{framed}
\usepackage[table]{xcolor}
\usepackage{graphicx}
\usepackage{booktabs}
\usepackage{enumitem}
\usepackage{microtype}
\usepackage[shortcuts]{extdash}

\newcommand{\popSize}[0]{803}

\begin{document}
\sloppy

\title{Hashing It Out: A Survey of Programmers' Cannabis \texorpdfstring{\\}{}Usage, Perception, and Motivation} 

\author{Madeline Endres}
\email{endremad@umich.edu}
\affiliation{%
  \institution{University of Michigan}
  \city{Ann Arbor}
  \state{Michigan}
  \country{USA}
}
\author{Kevin Boehnke}
\email{kboehnke@umich.edu}
\affiliation{%
  \institution{University of Michigan}
  \city{Ann Arbor}
  \state{Michigan}
  \country{USA}
}
\author{Westley Weimer}
\email{weimerw@umich.edu}
\affiliation{%
  \institution{University of Michigan}
  \city{Ann Arbor}
  \state{Michigan}
  \country{USA}
}

\renewcommand{\shorttitle}{Hashing It Out: A Survey of Programmers' Cannabis Usage, Perception, and Motivation}

\begin{abstract}
Cannabis is one of the most common mind-altering substances. 
It is used both medicinally and recreationally and is 
enmeshed in a complex and changing legal landscape.
Anecdotal evidence suggests that some software developers may use cannabis to aid some programming tasks. At the same time, anti-drug policies and tests remain common in many software engineering environments, sometimes leading to hiring shortages for certain jobs. 
Despite these connections, little is actually known about the prevalence of, and motivation for, cannabis use while programming. In this paper, we report the results of the first large-scale survey of cannabis use by programmers.
We report findings about  \popSize{} developers' 
(including 450 full-time programmers')
cannabis \textbf{usage prevalence}, \textbf{perceptions}, and \textbf{motivations}. For example,
we find that some programmers \emph{do} regularly use cannabis while programming: 35\% of our sample has tried programming while using cannabis, and 18\% currently do so at least once a month. Furthermore, this cannabis usage is primarily motivated by a perceived enhancement to certain software development skills (such as brainstorming or getting into a programming zone) rather than medicinal reasons (such as pain relief). Finally, we find that cannabis use while programming occurs at similar rates for programming employees, managers, and students despite differences in cannabis perceptions and visibility. Our results have implications for programming job drug policies and motivate future research into cannabis use while programming.
\end{abstract}




\maketitle

\section{Introduction}
\label{sec:intro}

\textit{Cannabis sativa} (hereafter, cannabis) is the world's most commonly used illicit substance, used by more than 192 million people in 2018~\cite{UNPressReport}.
The global cannabis market in 2020 was estimated at 20.5 billion USD, and is estimated to grow to 90.4 billion by 2026~\cite{businessWire}. 
Globally, cannabis legality is changing rapidly, with many countries (e.g., the United Kingdom, Colombia, and Malawi) legalizing medical cannabis, and a subset (e.g., Uruguay, Mexico, and Canada) also legalizing cannabis for recreational adult use.\footnote{\url{https://en.wikipedia.org/wiki/Legality_of_cannabis/}} In the US, for example, medical cannabis is allowed in 36 states, and 17 states have legalized cannabis\footnote{\url{https://medicalmarijuana.procon.org/legal-medical-marijuana-states-and-dc/}} for adult use despite its federal classification as a Schedule I drug, which criminalizes cannabis and defines it as having no accepted therapeutic value and a high abuse potential~\cite{NAP24625}. This classification has hampered research on its therapeutic effects~\cite{NAP24625}, and prohibition is contrary to popular opinion: 91\% of Americans believe cannabis should be legal for medical or recreational purposes~\cite{green2021pew}. Similarly, 81\% of Americans believe cannabis has at least one benefit~\cite{keyhani2018risks}. These benefits are mostly medical, 
but  16\% and 11\% cited improved creativity and focus, respectively.  

Despite legal concerns, anecdotes of cannabis use intersecting with software engineering abound.
Questions inquiring about cannabis's effects on programming are common on online forums such as Reddit,\footnote{\url{https://www.reddit.com/r/computerscience/comments/dbzp5v/how_many_of_you_found_that_smoking_weed_gets_you/}} 
Quora,\footnote{\url{https://www.quora.com/Are-there-any-pothead-programmers-Can-you-code-while-youre-high}} 
Hacker News,\footnote{\url{https://news.ycombinator.com/item?id=509614}} 
and Dev,\footnote{\url{https://dev.to/damcosset/coding-and-cannabis-3f8e}} often inspiring numerous conflicting answers. 
Similarly, popular tech-related media sites cover the topic, with one iTechPost article claiming that ``folks in the tech mecca that is Silicon Valley are clearly getting their fair share of medical (or otherwise) marijuana'', positing that cannabis may help with chronic pain associated with long hours of programming~\cite{iTechPost}. This claim aligns with epidemiological trends, as chronic pain is the most common reason for medical cannabis licensure in the US~\cite{boehnke2019qualifying}, and many people report that cannabis is useful for managing chronic pain symptoms~\cite{BOEHNKE2019830}. Congruent with popular opinion, other posts claim that cannabis products enhance programming through increasing focus~\cite{programmingInsider} or creativity~\cite{simpleProgrammer}. However, to the best of our knowledge, no study has investigated motivations of focus, creativity, or wellness for cannabis use while programming. 

Cannabis is often prohibited in the workplace, a policy frequently enforced through mandatory drug testing for metabolites of $\Delta$-9-tetrahydrocannabinol (THC), which causes intoxication. A 2018 report found 63\% of US organizations conduct drug screening, with two-thirds not accommodating medical cannabis use or lacking a medical cannabis policy~\cite{HRReport}. In software engineering workplaces, drug tests remain common: we find that 29\% of programmers have taken a drug test for a programming-related job (see Section~\ref{sec:cannabis_usage}). 

This prohibition of cannabis use in software engineering has contributed to a widely-reported hiring shortage for certain US government programming jobs~\cite{viceFBI2014, bbcFBI2014, wsjFBI2014}. In 2014, after struggling to meet hiring goals, then-director of the American Federal Bureau of Investigation James Comey stated that while he had ``to hire a great work force to compete with \dots cyber criminals[,] \dots some of those kids want to smoke weed on the way to the interview'', a behavior counter to the FBI's current policy that ``prohibits anyone working for it who has used cannabis in the past three years''~\cite{bbcFBI2014}.  

Despite such evidence highlighting the intersection of cannabis and programming, 
little empirical research has been conducted on cannabis use in software development. To our knowledge,  previous related publications either combine cannabis use with other illegal substances such as cocaine~\cite{hoffman1996drug}, show data about computer scientists as a sub-population in a much larger survey~\cite{mensch1988job}, and/or include programming only as part of a much broader sector such as engineering or business services~\cite{mensch1988job}. 
These previous studies do not sufficiently investigate the intersection of cannabis and programming for several reasons.  First, substances such as cocaine have drastically different physiological effects than cannabis and are used for different reasons: grouping such disparate substances may result in ineffective blanket drug policies and misleading conclusions. Second, studies that group jobs by sector often focus on physical consequences of cannabis usage (e.g., impairment while operating a machine) that are less relevant to software (where our participants place more of a focus on, e.g., creativity). 
Thus, it remains unclear if, when, or why people currently use cannabis while programming, let alone how job-type or company policy may play a role. Without a grounded understanding of this intersection, companies and hiring managers cannot effectively evaluate the utility of extant anti-cannabis policies. 
We present results from the first large-scale empirical survey of cannabis use in software engineering, reporting data from \popSize{} programmers, including 450 full-time developers (via an online, anonymized and confidential survey):  

\begin{itemize}[leftmargin=10pt,topsep=2pt]
    \item We find that some developers (18$\%$ of our sample) use cannabis while programming at least once a month, with many even choosing to use it for work-related tasks.
    \item We find that cannabis use while programming is more commonly motivated by perceived programming-related skill enhancement than by medical reasons. This aligns with perceptions among a subset of students and younger people that cannabis use may enhance creativity or cognitive performance~\cite{Andreas2016, anthenien2021cannabis}.
    \item We find that cannabis use while programming occurs at similar rates for programming employees, managers, and students despite differences in perception of cannabis approval level and cannabis visibility between the three groups. 
    \item Despite contrary anecdotal reports, we find that anti-cannabis policies and screening remain common for programming-related jobs, with 29$\%$ of respondents reporting having taken a drug test for a programming-related job. 
\end{itemize}
We close with a discussion of the implications of our results on software company drug policies and future research directions. 

\section{Background and Related Work}
\label{sec:relatedwork}

 We provide context about the legality and uses of cannabis, focusing on cannabis use the United States as we scoped our study to US-based developers. However, we note that recent cannabis legality shifts in the US align with global trends. We close with a discussion of prior work at the intersection of programming and cannabis.

\textbf{General Cannabis Background}---In the past century, cannabis legality has shifted dramatically in the US. While many cannabis preparations were listed in the US pharmacopoeia in the early 20th century, cannabis was officially criminalized under the Controlled Substances Act in 1970. 
Since 1996, however, 36 states have legalized cannabis for medical purposes, and 17 for adult use. The policy whiplash has been largely driven by politics and popular opinion rather than science, with the re-emergence of cannabis occurring partially due to: 1) acknowledgement of the cruel excesses and ineffectiveness of the War on Drugs~\cite{woodWar2009, edwards2020tale};
2) increasing acknowledgement of cannabis's therapeutic benefits~\cite{NAP24625};
and 3) the ongoing opioid crisis, which has driven home the need for alternative, less harmful pain medicines~\cite{choo2016Opiods}. 
In conjunction with liberalizing policies, past-year cannabis use has increased among Americans 12 or older: 11\% in 2002 to 17.5\% in 2019~\cite{NSDUH}.  

Cannabis is used for many reasons, including medical (e.g., pain, nausea) and recreational (e.g., social or perceptual enhancement, altered consciousness) purposes~\cite{BOEHNKE20191362, lee2009development}. 
This wide variety of uses is enabled by the pharmacopeia of compounds in cannabis, which include more than 100 active cannabis-derived compounds (cannabinoids) such as cannabidiol (CBD) and $\Delta$-9-tetrahydrocannabinol (THC) as well as terpenes and flavanoids, which are responsible for taste and odor~\cite{NAP24625}.
The most common medical reason for cannabis use is chronic pain, demonstrated by patient surveys~\cite{baron2018patterns, reiman2017cannabis} 
and registry data for medical cannabis patients~\cite{boehnke2019qualifying}. Within broad categories of medical vs. recreational use, however, there is subtle shading of motivations. Lee \emph{et al.} developed the comprehensive marijuana motives questionnaire, in which they identified factors associated with cannabis use, including enjoyment, coping, experimentation, altered perception, sleep/rest, and availability~\cite{lee2009development}. These factors are quite broad, and may encompass recreational, medical or other use patterns such as for creative work or performance enhancement. 

\textbf{Cannabis and Programming}---Within this complex environment, cannabis interacts with software development. For instance, cannabis use in Silicon Valley appears to be high, with area dispensaries reporting that around 40\% of their clientele are tech workers~\cite{bloomberg2013Silicon}. Similarly, a qualitative study of coding bootcamps identified ``lots and lots of [weed]'' as one key element of support~\cite{byrd2019between}. 

No previous empirical studies have directly investigated the intersection of software and cannabis. Reports either combine cannabis with other illegal substances or include programmers only as a sub-population~\cite{hoffman1996drug, mensch1988job}. For instance, one 
study has a table of the percentage of workers in various fields reporting illegal drug use~\cite[Tab~3.5, p.~26]{hoffman1996drug}. Programmers appear in the table, with 10.4\% reporting usage in the last year. However, this is the only mention of programmers, cannabis is not separated from other  drugs, and the data is from the early 1990s, well before cannabis was legalized in numerous states. Thus, questions of cannabis prevalence, usage motivations, and culture in software engineering remain unanswered. 

However, cannabis use has been studied in the context of creative problem solving, a key component of many software engineering tasks~\cite{Groeneveld2020Sigcse, Groeneveld2021ICSE, maiden2006Creative}. Generally, cannabis is seen as a creativity enhancer: in one study, 54\% of participants believed it increased their creativity~\cite{green2003being}. Similarly, Steve Jobs, who was regularly using cannabis when he founded Apple, stated that ``the best way [he] would describe the effect of the marijuana \dots is that it would make [him] relaxed and creative''~\cite{wiredJobs}. From a cognitive perspective, the link between creativity and cannabis use is more tenuous. LaFrance \emph{et al.} found that cannabis users were more creative than non-cannabis users, however, this difference may be explained by underlying personality~\cite{lafrance2017inspired}. On the other hand, Kowal \emph{et al.} found that highly potent cannabis can actually \textit{impair} divergent thought, a type of creative thinking~\cite{kowal2015cannabis}. These conflicting results linking cannabis and creative problem solving may motivate future observational studies of cannabis and programming. 

Apart from creativity, cannabis has a range of other cognitive effects which could impact software development; long-term cannabis use can impair memory and attention control~\cite{Kroon2021TheShort}. However, even for general cognitive effects, the current scientific literature is often insufficient; evidence regarding cannabis's long-term effects on skills used in software engineering such as decision making, motor ability, and working memory is inconsistent (see Kroon \emph{et al.}~\cite{Kroon2021TheShort} for a review of the current research on cannabis and cognition). 

\textbf{Programming and Other Mind-Altering Substances}--While the intersection between programming and cannabis remains unexplored, there exists research into connections with other mind-altering substances such as alcohol and LSD. One study of IT professionals in India linked high levels of work-related stress to increased rates of alcohol abuse and depression~\cite{darshan2013study}. Jarosz \emph{et al.} linked alcohol and creativity, finding that intoxicated subjects completed more creative problem solving tasks than sober counterparts~\cite{Jarosz2012Uncorking}. As for LSD, psychedelics have historical cultural connections with software development~\cite{walsh2011drugs, markoff2005dormouse}: LSD was commonly reported by many early PC and internet developers as a source of innovation and creativity~\cite{walsh2011drugs}, a phenomenon likely connected to 1960s counterculture~\cite{markoff2005dormouse}. However, despite preliminary evidence that micro-dosing on psychedelics may increase creativity~\cite{prochazkova2018exploring}, connections remain cultural and anecdotal rather than causal, further motivating future studies of mind-altering substances and programming.


\vspace{-1pt}
\section{Survey Design and Methodology}
\label{sec:design}

To understand programmers' cannabis usage, perception, and motivation, we conducted an online survey of \popSize{} programmers. We desired a design that was \emph{time-efficient} (allowing for a large number of responses), \emph{low risk} (free of legal or employment-related consequences), and structured to permit \emph{statistical} comparisons (within sub-populations such as employment status, age, and gender). 

Thus, we designed our survey to take under 30 minutes, to be anonymous and confidential, and to use best practices from drug survey construction in other fields, including adapting questions from previously-validated surveys when possible.
We scoped this study to United-States-based developers. To obtain a diverse array of responses, we recruited participants from several sources including GitHub, social media, and the University of Michigan (a large public American university). We did not recruit participants directly from software companies to avoid risks of work retaliation. Recruitment materials indicated that previous cannabis use was \textit{not} necessary to participate.

We now describe the construction (Section~\ref{sec:survey_construction}), ethical considerations (Section~\ref{sec:ethics}), and distribution (Section~\ref{sec:survey_distribution}) of our survey. Our replication package is available online with our full survey, recruitment materials, and IRB protocol.\footnote{Replication package with survey instruments, data analysis scripts, and our IRB protocol: \url{https://github.com/CelloCorgi/HashingItOut_ICSE2022}.}

\subsection{Survey Construction}
\label{sec:survey_construction}

Our survey included questions on demographics, programming background, cannabis attitudes, and cannabis usage. 
For cannabis sections, we asked questions about cannabis in general and in relation to programming. 
We employed display logic as appropriate to minimize exposure to irrelevant questions. 
When possible, we based questions on previous studies about cannabis~\cite{green2021pew, NSDUH, cuttler2017measuring} or software development~\cite{tilley2005} to allow comparisons with prior work. 

\textbf{Demographics}--Querying age, gender, and employment status, our demographics questions were adapted from those asked by Boehnke \emph{et al.} in their recent survey of cannabidiol use for fibromyalgia~\cite{BOEHNKE2021556}. To help ensure participant safety, we did \emph{not} collect identifying information such as names or IP addresses. As a result, we included additional validity checks as described in Section~\ref{sec:survey_distribution}. 

\textbf{Programming background}--Participants reported how long they had programmed, their programming education, and their programming-related job history.
We also asked participants with programming jobs if they self-identified as a manager, and we asked participants to indicate how often they conducted various software-engineering related tasks such as brainstorming and requirements elicitation, adapted from those used by Tilley \emph{et al.}~\cite[Sec.~2]{tilley2005}.

\textbf{Cannabis attitudes}---We asked questions regarding attitudes toward cannabis, both in general and also in programming-specific contexts. 
To gauge general cannabis attitudes, we asked questions about cannabis legalization and perceived risk drawn from previous American national surveys from the Pew Research center~\cite{green2021pew} and the National Survey on Drug Use and Health~\cite{CARLINER201713, NSDUH}. For programming-specific cannabis attitudes, we adapted questions previously used to assess cannabis attitudes or perceptions from other contexts, e.g., from a study of high school seniors' disapproval about individuals carrying out cannabis-related activities~\cite{bachman1988explaining}.

\textbf{Cannabis usage}--We asked questions regarding general and programming-specific cannabis use. For \textit{general cannabis use}, we asked questions from the \textit{Daily Sessions, Frequency, Age of Onset, and Quantity of Cannabis Use Inventory} (DFAQ-CU)~\cite{cuttler2017measuring}, a validated measure of cannabis use behavior and history. We included questions that quantify current use frequency, past periods of heavy use, medicinal vs. recreational cannabis, and the percentage of the time participants use cannabis products with THC (see Section~\ref{sec:relatedwork}). 
We also asked how COVID-19 affected cannabis usage patterns. To measure cannabis use \textit{while programming}, we adapted questions from the DFAQ-CU, adding the phrase ``while programming, coding, or completing any other software engineering-related task?''. We also asked for which types of programming projects or tasks are participants likely to use cannabis, and we asked participants about their motivation(s) for using cannabis with choices reflecting those we observed in anecdotal online posts as well as a free text option. 

\subsection{Ethical Considerations}
\label{sec:ethics}

In the US (the source of our population), cannabis remains illegal at a federal level and in many states, so use can result in fines or incarceration. Further, cannabis is often explicitly prohibited by corporate policy, potentially resulting in employee termination. As such, we worked closely with our IRB to minimize legal risks and ensure participants felt comfortable answering honestly. First, we made our survey anonymous and confidential: we did not collect names or IP addresses, even though doing so necessitates additional data-quality checks. Second, all participants gave informed consent, and all questions were optional. Third, we focused our recruitment of professional developers on open source projects rather than through software engineering company contacts to avoid the risk of work place retaliation or coercion. Fourth, we collected emails for our optional incentive on a separate platform where they could not be connected back to survey responses. Finally, we are unable to publish our full data set: although anonymous, it contains demographic data which may inadvertently identify participants. 

\subsection{Survey Distribution}
\label{sec:survey_distribution}


\textbf{Survey platform}---When choosing a survey platform, we wanted to ensure the data we collected was anonymous and confidential while also providing data quality assurance. Through consultation with our IRB, we used the Qualtrics XM Platform
which enabled anonymous and confidential collection as well as data-quality options such as preventing multiple submissions and bot detection.


As mentioned in Section~\ref{sec:design}, we scoped this initial study of cannabis use and programming to United-States-based developers due to the high variance of cannabis laws and cultures worldwide. All recruitment materials stated that the survey was optional, anonymous, and confidential. To help mitigate participant self-selection bias, we also made it clear that prior cannabis use was not required to participate. Finally, survey participation was encouraged through an optional drawing for one of five \$100 awards. All data was collected between March 31 and May 2, 2021. 

\textbf{Survey recruitment}---To encourage diverse responses, we recruited from several populations: open-source GitHub developers, current and former computing students at the a large public university, and social media users. 

For participants recruited from GitHub, we used GitHub’s REST API to obtain the top 1000 developers and top 100 repositories associated with each of 25 ``popular'' programming languages (as identified in the GitHub interface) and 8 additional common languages such as MATLAB and R. For each repository, we pulled the profiles of the top 25 contributors. We filtered for profiles with a public email, resulting in 31,259 potentials. Using regular expressions and manual review, we identified 7,372 with a US location. Eliminating an additional 1,613 using DNS verification, we sent 5,759 emails, of which 36 failed, and received 440 valid responses (7.7\%), a rate similar to previous studies of open-source developers~\cite{shull2007guide, HuangICSE2021}. This use of GitHub profiles for research is permitted by the terms of service.\footnote{\url{https://docs.github.com/en/github/site-policy/github-acceptable-use-policies}, see Section 6.}

As for the university-recruited participants, we sent 5,638 emails to all current and former undergraduates who took a programming course for CS-majors (e.g., CS2 or CS3) at the University of Michigan between Fall 2018 and Fall 2019, receiving 283 responses (5.0\%, 12 failed). As this study was conducted in 2021, this strategy recruited a mix of more senior CS undergraduates and young industrial developers rather than only current students. We also emailed CS graduate students for 56 responses. Finally, we posted the survey on Twitter, yielding 24 responses. While our study is not a random sample, we note that convenience samples are common in the cannabis and hidden populations literature~\cite{miller2010using} (see Section~\ref{sec:limitations}).

\textbf{Survey data validation}---In total, 1045 participants started our survey. To ensure we analyzed only high-quality data, we implemented several post-collection checks. First, 236 partial responses were removed. While participants could skip individual questions due to the sensitivity of the survey topic, valid responses must have answered at least: 1) if they had programmed; 2) if they had used cannabis, and; 3) if they had used cannabis while programming. To mitigate rushing through the survey, we removed responses completed faster than 1.5 standard deviations below the median. Finally, we checked for consistency between reported age, years of programming experience, professional programming experience, and cannabis use. 
Combined, our completion time threshold and consistency checks eliminated 6 participants, leaving \popSize{} valid participants for analysis.  

\section{Population Contextualization}
\label{sec:pop_context}
We now present indirect evidence that our participants, while not a random sample, are similar in many ways to previous random samples or studies. A true random sample would not have been ethically permitted, but we gain confidence in our results' generalizability by contextualizing participants' gender, age, and employment. 

\begin{table}[t]
\small
\begin{tabular}{@{}lrrrr@{}}
\toprule

Population             & Overall            & GitHub             & University         &Social Media \\ \midrule
                         
All Responses         & 803                & 440                & 339                & 24           \\ 
Man                      & 666 (83\%)         & 403 (91\%)         & 243 (72\%)         & 20 (83\%)    \\
Woman                    & 112 (14\%)         & 21 (5\%)           & 87 (26\%)          & 4 (17\%)     \\
Non-binary               & 20 (2\%)           & 12 (3\%)           & 8 (2\%)            & 0 (0\%)      \\
No Answer                & 5 (\textless{}1\%) & 4 (\textless{}1\%) & 1 (\textless{}1\%) & 0 (0\%)      \\ 
Median age               & 26                 & 34                 & 21                 & 29           \\
Average age              & 29.2               & 34.9               & 21.7               & 31.0         \\
Min /Max age             & 15 / 70            & 15 / 70            & 18 / 61            & 21 / 52      \\ \midrule
\multicolumn{5}{l}{\textit{Employment Status (could select multiple)}}                    \\ \midrule
Full-time CS job         & 450                & 357                & 75                 & 17           \\
Student                  & 290                & 34                 & 251                & 5            \\
Unemployed               & 54                 & 19                 & 34                 & 1            \\
Part-time CS job         & 50                 & 18                 & 30                 & 2            \\ \midrule
\multicolumn{5}{l}{\textit{Programming-Related Job Title (could select multiple)}}             \\ \midrule
Software Engin.          & 311                & 232                & 70                 & 9            \\
Developer                & 270                & 190                & 70                 & 10           \\
Systems Engin.           & 72                 & 54                 & 15                 & 3            \\
CS Researcher            & 53                 & 23                 & 25                 & 5            \\
CS Instructor            & 49                 & 29                 & 20                 & 0            \\
Data Scientist           & 49                 & 25                 & 24                 & 0            \\
Product Manager          & 22                 & 17                 & 4                  & 1            \\
IT                       & 21                 & 16                 & 5                  & 0            \\ \bottomrule
\end{tabular}
\caption{Demographics overview of our survey population broken down by recruitment pool. We note that participants in the `University' column are those recruited from university emails. However, by the time of recruitment, many such participants had already graduated and thus have full-time CS jobs. Finally, we note that for job titles, participants could, and commonly did, select multiple descriptors. Thus, numbers in this section may add to more than the number of participants with jobs (e.g., many selected both Software Engineer and Developer).  } 
\label{tab:demogrpahics}
\end{table}

Table~\ref{tab:demogrpahics} overviews the gender, age, and programming-related employment of our study population by recruitment pool (e.g., how they were contacted to participate in this study). The majority (83\%) of our population are men. This percentage is higher in those recruited from GitHub (91\%) than those from the university (72\%) and social media (83\%). While this gender gap is large, it is similar to what we would expect of our sample population as a whole. For instance, Vasilescu \emph{et al.} found that, of public GitHub profiles with ascertainable gender, 91\% were men~\cite{DBLP:conf/chi/VasilescuPRBSDF15}, the same percentage observed in our sample. We see a similar correspondence with university-recruited participants: 26\% are women, similar to the 24\% of our CS undergraduate population overall. Because they are similar to those of our recruitment populations, our observed gender ratios give confidence in the generalizability of our results even though we could not collect a random sample of all developers.

The ages of participants also align with our sampled population. Our participants range in age from 15 to 70, with an average age of 29.2. As expected, GitHub-recruited developers were generally older than university-recruited participants, with an average age of 34.9 compared to 21.7. This age of our GitHub participants is comparable to the 30 reported in a previous study of open source developers~\cite{lakhani2003hackers}. Similarly, the average age of our university-recruited participants matches a typical US senior undergraduate.

As for employment, the majority (56\%) of our sample are currently full-time employees at a programming-related job while an additional 6\% are part-time and 36\% are students. Of those that currently have a programming-related job, we observe a wide range of reported  titles. While the most common titles were software engineer and developer (30\% and 34\% of our sample respectively), a significant number of participants identified as a systems engineer, computer science researcher, computer science instructor, data scientist, project manager, or information technician (between 2--10\% of our sample for each).\footnote{Participants could select multiple job titles. We do note that 27 respondents (around 3\% of  our sample) self-reported as only CS Instructors. As suggested by our reviewers, we conducted an additional sub-population analysis removing these participants from our sample. This removal resulted in no changes to our overall significance results and analysis conclusions. For example, the number of developers who have tried cannabis while programming is 35.4\% with educators removed and 34.8\% with educators included. We include the full results of this additional analysis in our replication package.} This wide array of jobs indicates that our sample contains a diverse sample of programmers in various fields.

As for self-identified students, the majority came from university-recruited emails. Due to survey time constraints, we did not include a functional test of programming ability for students. However, all university-recruited participants had taken a programming course for CS-majors 1--3 years before participating. This gap was intentional: it resulted in recruiting young computing professionals rather than only students. Since some in this set had graduated, the ``University'' descriptor in Table~\ref{tab:demogrpahics} refers to the email list source rather than current enrollment. To verify that students were indeed more advanced, we asked about general and professional programming experience. For general experience, results match expectations for more advanced undergraduates and graduates: both the first quartile and median student participant reported 3--5 years of experience while the third quartile reported 6-10 years. 67\% of students also reported professional programming experience. Of these students, the median was 1--3 years, with some students reporting 6--10. This high level of professional experience may reflect the 21\% who were graduate students.

Overall, the gender and age of our participants aligned with their populations. Also, even though we only collect data from university emails and open-source users, our sample's high percentage of professional developers and wide array of programming-related jobs indicates that our sample contains diverse types of programmers. We thus gain confidence in the generalizability of our findings.

\section{Research Questions}

We organize our analysis around the following questions:
\begin{itemize}[leftmargin=15pt,topsep=2pt]
    \item \textit{RQ1---Usage:} Do programmers use cannabis while programming? If so, how often?
    \item \textit{RQ2---Context:} In what contexts do programmers use cannabis?
    \item \textit{RQ3---Motivation:} Why do programmers use cannabis? 
    \item \textit{RQ4---Perception:} How do opinions of programming cannabis use vary between managers, employees, and students? 
    
\end{itemize}

\textbf{Statistical Methods---}Our analysis was conducted in a Python Jupyter Notebook
using \texttt{Pandas}~\cite{pandas}. For our statistical analyses, we primarily used \texttt{SciPy}~\cite{2020SciPy-NMeth} and \texttt{Statsmodels}~\cite{seabold2010statsmodels}.
When testing the significance of a difference between continuous variables (e.g., age) or Likert scores (e.g., 5-point scale) of two independent sub-populations, we use the Student's $t$-test. While Likert scores are ordered categorical variables, previous research shows that with large samples, parametric tests are sufficiently robust for analyzing Likert data even though it is ordinal and normality cannot be assumed~\cite{likertStatistics}. Thus, the Student's $t$-test is best statistical practice. For testing the significance of a difference between two binary variables, we use the $n$-1 $\chi^2$-test (i.e., the proportions $z$-test)~\cite{campbell2007chi}.

We consider results significant if $p<0.05$. As this is a large survey study, we investigate multiple research questions and conduct multiple statistical tests. To avoid fishing and $p$-hacking, we defined our primary research direction when designing the survey and we report the results for all initial research questions and analyses. Within each research question, we also correct for multiple comparisons using a Benjamini-Hochberg False Discovery Threshold of $q=0.05$: unless otherwise noted, all significant results pass this multiple comparisons threshold. The majority of our findings produced $p$-values well below 0.0001, increasing confidence.

\subsection{RQ1: Cannabis Usage While Programming}
\label{sec:cannabis_usage}

We first investigate if and how often programmers in our sample use cannabis while coding. We analyze usage trends in our sample overall and by gender, age, and recruitment pool. 

\textbf{Overall cannabis usage while programming}---Overall, we find that 35\% (280/\popSize{}) of our participants have tried cannabis while programming or completing another software engineering-related task, approximately half of those who tried cannabis in general (69\% = 557/\popSize{}). Of those that have used cannabis while programming, 73\% (205/280, 26\% of our population overall) used cannabis while programming in the last year. While not a perfect comparison, we observe higher cannabis use than that in recent national surveys: 35\% of Americans ages 18--25 and 15\% of those 26 and older report using cannabis in the last year~\cite{NSDUH} compared to 54\% of our sample. However, considering our population (and open source developers in general) skews young and male, higher reported use is expected. 

\begin{figure}[t]
    \centering
    \includegraphics[width=\linewidth]{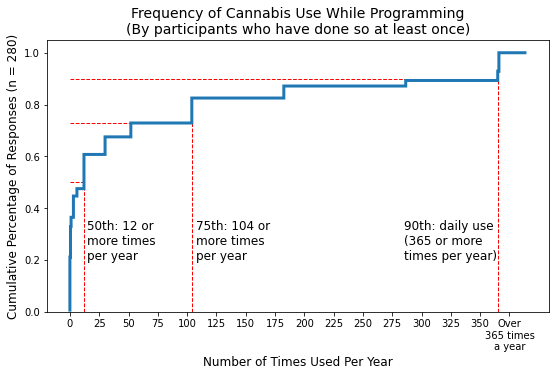}
    \caption{``Which of the following best captures the average frequency you currently use cannabis while 
    programming, coding, or completing any other software engineering task?'' 
    responses by programmers who have used cannabis while programming at least once. Converted from options such as ``1 time a week'' to times used per year for clarity. Dotted lines show the approximate 50th, 75th, and 90th percentiles of use frequency. For example, programmers who currently use cannabis while programming at least 104 times per year are in the 75th percentile of use frequency.}
    \label{fig:useFrequency}
\end{figure}

\textbf{Cannabis usage frequency}---We also investigate \emph{how often} participants currently use cannabis while programming. In the last year, 53\% (147/280, 18\% of our full sample) reported using cannabis while programming at least 12 times (monthly). Furthermore, 27\% (76/280) reported using while programming at least twice a week (100 times per year) and 11\% (30/280, 4\% of our sample as a whole) reported using on a near daily basis. While those frequencies speak to \emph{current} usage over the last year, these trends also occur over a longer term: 46\% of our sample of cannabis-using programmers (128/280) also report that they have, at a point in the past, regularly used cannabis while programming (2 or more times per month for at least one six-month span). Our results regarding cannabis use frequency while programming are visualized in Figure~\ref{fig:useFrequency}.

These findings give the first formal insight into the prevalence of cannabis in programming communities, and have important implications for drug tests in software engineering. 
Urine-based drug tests detect cannabis up to 30 days after use~\cite{hadland2016objective}, much longer than the interval between cannabis sessions reported by many developers. Additionally, programmers' cannabis products typically contain THC, the compound detected by most drug tests~\cite{hadland2016objective}: on average, developers reported that 87\% of their cannabis had THC, with the median reporting 100\% of products included the compound. Simultaneously, we found that drug tests remain common for software: 29\% of our sample reported that they had taken a drug test for a programming-related job. Thus, cannabis-using developers may avoid applying to jobs with drug tests, limiting application pools.

\textbf{Cannabis use demographic context}---We further contextualize our results by investigating variance by gender, age, and recruitment pool. Regarding gender, we do not observe a significant difference between men and women in the percent of participants who have tried cannabis (70\% vs. 69\%, $p=0.87$). However, we do find that men are more likely than women to have tried cannabis while programming (36\% vs. 25\%, $p<0.0001$). This aligns with surveys of general populations not limited to programmers which find that men use cannabis more frequently than women~\cite{cuttler2016sex}. Even though our sample size is smaller, we also observe that non-binary and transgender participants (not broken down in Table~\ref{tab:demogrpahics} for space) are significantly more likely to have tried cannabis while programming (57\% vs. 34\%, $p=0.01$) than the rest of the population.

We also find a small but significant positive correlation between age and current frequency of cannabis use while programming ($r=0.21$, $p=0.003$):\footnote{Calculated using Spearman's $r$ which detects all monotonic relationships (as opposed to Pearson's $r$ which detects only linear relationships).} of those who currently use cannabis while programming, older programmers tend to use it slightly more often. 
When plotting this correlation, we observe the bulk of this increase can be attributed to increases in usage likelihood before the age of 35. After that, usage frequency appears to level off or drop slightly. 

We also examine our two main recruitment pools: GitHub and university. 
GitHub participants were not significantly more likely to have tried cannabis (71\% vs. 65\%, $p=0.065$). However, they were more likely to have used cannabis while programming (39\% vs. 28\%, $p=0.001$). This aligns with observed demographics as GitHub participants are more likely to be men and tend to be older. 
However, it also may indicate that cannabis use is particularly prevalent in open source communities, an observation that motivates future qualitative investigation of open-source cannabis culture. 

\begin{framed}{
    \noindent Over one-third of our sample have used cannabis while completing a programming or software engineering-related task, of which half currently use cannabis at least once or more a month. Programmers typically use cannabis products that contain THC, and 11\% of programmers who have used cannabis while programming report currently doing so on a near daily basis, behaviors very likely to be detected by most drug-test related policies. We find that cannabis use while programming is particularly common for non-binary or transgender participants (57\%) and participants recruited from GitHub (39\%).  
}
\end{framed}

\subsection{RQ2: Cannabis Use Programming Context}
\label{sec:prog_context}

We now investigate in what programming-related contexts programmers use cannabis, including both high-level project qualities (e.g., personal projects vs. work-related projects) and also software-engineering task types (e.g., refactoring, debugging, or requirements elicitation). We also analyze the impact of remote work.

\textbf{In which programming contexts is cannabis used?}---We first investigate which programming project types (e.g., personal or work projects) developers are most likely to choose to complete while using cannabis. 
We provide our full results in Table~\ref{tab:project_types}, but we emphasize our result that 95 participants (34\% of cannabis using programmers and 12\% of our population overall) sometimes use cannabis for work-related tasks. While we anticipated our finding that personal programming projects would be the most common project type completed while using cannabis, we hypothesized that the percentage using cannabis for work-related projects would be lower than observed. This indicates that cannabis routinely interacts with professional software engineering environments, underlining the potential impact of corporate drug policies and motivating future studies of cannabis in software-engineering.

\begin{table}[t]
\begin{tabular}{@{}lrr@{}}
\toprule
Project Type                        & Number & Percent   \\ \midrule
Personal programming projects       & 175    & 63.0 \\
Non-urgent programming tasks        & 133    & 47.8 \\
Work-related programming tasks      & 95     & 34.2 \\
School-related programming tasks    & 76     & 27.3 \\
Deadline-critical programming tasks & 25     & 9.0  \\ \bottomrule
\end{tabular}
\caption{Project types for which participants sometimes choose to use cannabis while programming. Percentages are out of the 280 who have used cannabis while programming.}
\label{tab:project_types}
\end{table}

\begin{figure}[t]
    \centering
    \includegraphics[width=\linewidth]{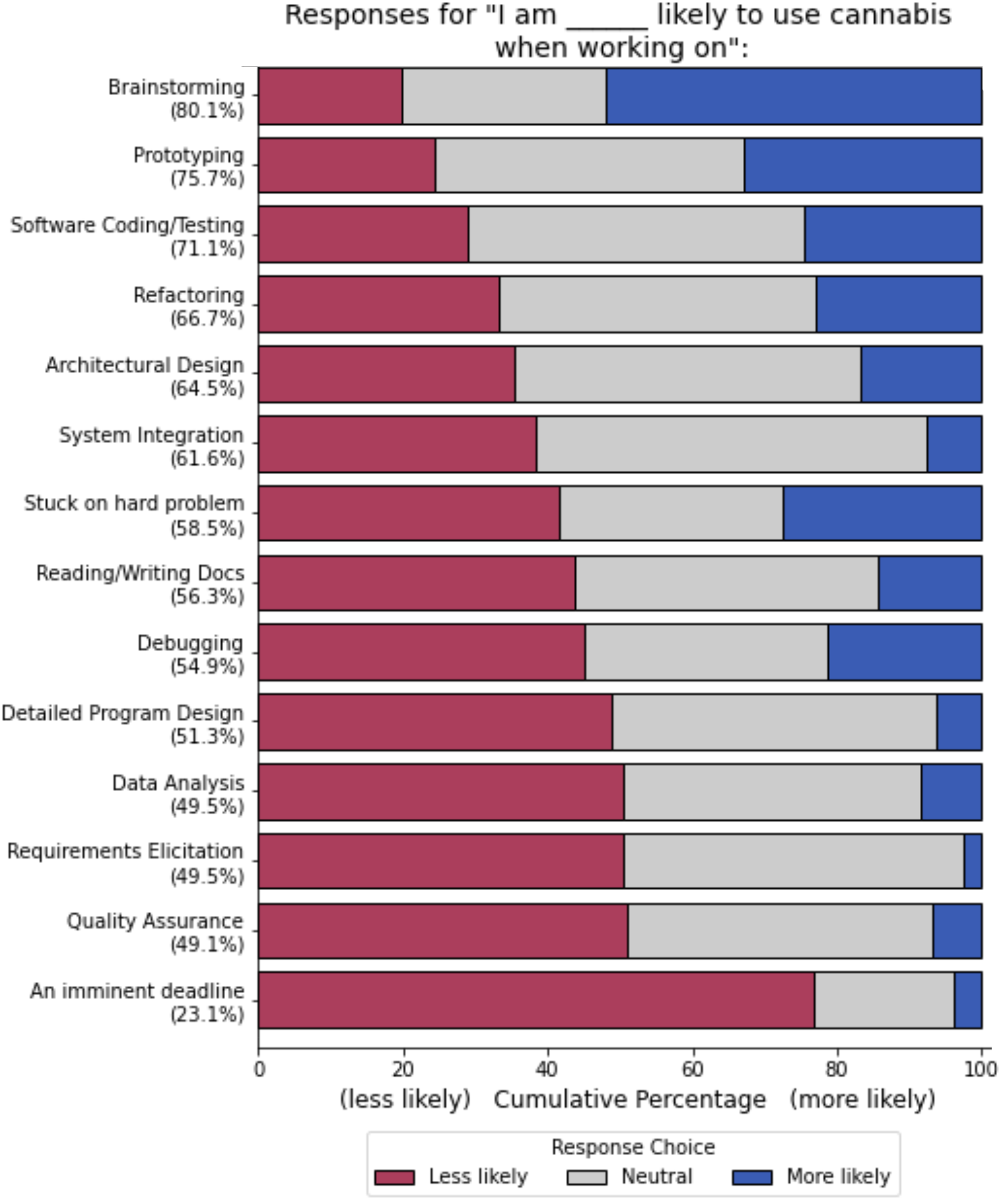}
    \caption{Chart of participant cannabis use likelihood while completing various common programming-related tasks on a 3-point scale. Tasks ordered by combined percentage of more likely and neutral to use cannabis. Percentages are the sum of more likely and neutral responses for each task.
    }
    \label{fig:whenUseCannabis}
\end{figure}

\textbf{For which software tasks do programmers use cannabis?}---We now investigate how likely participants are to use cannabis while completing common software engineering tasks adapted from Tilley \emph{et al.}~\cite{tilley2005}. 
Our full results are in Figure~\ref{fig:whenUseCannabis}: programmers reported a higher likelihood of using cannabis while brainstorming or prototyping and a lower likelihood of using cannabis while performing quality assurance, requirements elicitation, or tasks with an imminent deadline. 
These results indicate that developers may self-regulate cannabis use for when it is most beneficial (i.e., for creative, open-ended tasks) while avoiding use for time- or safety-critical tasks. We note that participants who are unable to self-regulate cannabis use (e.g., are dependant on cannabis) may be unlikely to admit so in our survey. 
Even so, our results call into question the usefulness of blanket anti-cannabis policies. We investigate the motivations of these choices further in Section~\ref{sec:rq_motivation}.

\textbf{Cannabis use and remote work}---Many developers work remotely. Also, the COVID-19 pandemic was ongoing during recruitment. We find that 52\% (145/280) of cannabis-using programmers report they are somewhat or a lot more likely to use cannabis for work-related tasks when working from home compared to only 5\% (13/280) who report that they are somewhat or a lot less likely to do so ($p<0.0001$). Similarly, 29\% (82/280)  report increased programming cannabis use since the onset of COVID-19 compared to only 10\% (27/280) who report a decrease ($p<0.0001$), a result in line with other populations such as medicinal cannabis users~\cite{boehnke2021medication}. This indicates that workplace culture, environment, and policies can tangibly effect the frequency of cannabis use while programming. 

\begin{framed}{
    \noindent Developers most commonly choose to  use cannabis during personal programming projects (63\%). However, over a third of cannabis-using programmers also sometimes choose to use cannabis during work-related tasks, use that is more common during remote work. Programmers also self-regulate when they use cannabis: cannabis is more likely during creative open-ended software tasks vs. time- or safety- critical tasks. 
}
\end{framed}
\subsection{RQ3: Cannabis Use Motivation}
\label{sec:rq_motivation}

Having established that some developers regularly use cannabis while completing both personal and work-related programming projects, we now investigate \emph{why} programmers use cannabis. 
Understanding why developers use cannabis is important because it can help inform company drug policies and developer support. 

\textbf{Overall Motivation Results}---Our results on cannabis use motivations are reported in Table~\ref{tab:whyUse}. Overall, we found that programmers were more likely to report enjoyment or programming enhancement motivations than wellness motivations: the most common reasons were ``to make programming-related tasks more enjoyable'' (61\%) and ``to think of more creative programming solutions'' (53\%). In fact, all programming enhancement reasons were selected by at least 30\% of respondents. On the other hand, general wellness related reasons (such as mitigating pain and anxiety) were all cited by less than 30\% of respondents. Thus, while wellness does motivate some cannabis use while programming, it is not the most common motivation. This result is further corroborated by only 19\% (54/280) of cannabis using programmers indicating that they have a physician’s recommendation to use cannabis medicinally. Additionally, of those that have such a recommendation, two thirds report using cannabis for both medicinal and recreational reasons. This is important because it indicates that any cannabis policy should consider medicinal, recreational and performance enhancing marijuana use. 

We also investigated cannabis-use motivations by population pool (i.e., GitHub-recruited vs. University-recruited participants)---while percentages varied slightly,
the top four rationales were the same regardless of recruitment pool. Wellness responses were also consistently below programming-enhancement motivations.

\textbf{Additional Qualitative Responses}---While we leave a formal qualitative analysis to future work, we also observe the emphasis of programming-enhancement-related reasons for cannabis use while programming in the textual free-response section. For instance, one participant said that when using cannabis while programming, they are ``able to better connect ideas and think about things on a broader level, which typically leads to more well-rounded solutions.'' Similarly, another participant stated that cannabis use while programming helped brainstorming, allowing them ``to organize [their] thoughts better and keep them separate, which helps [them] follow threads further and come up with new paths to follow''.

Beyond cannabis's apparent usefulness during programming itself, participants also cite its usefulness during adjacent tasks. For example, one participant said that they used cannabis ``to stay awake/focused when \dots grading 70 programming assignments. If [I] have a deadline and need to binge work for many hours, it is easier to keep going if I periodically get high. When your life is mostly work, cannabis is something that makes it bearable.''  Finally, some participants indicated reasons other than enjoyment, wellness, or programming-enhancement for using cannabis while programming. For instance, some participants indicated that cannabis use while programming was a coincidence of regular cannabis use while a few others said they tried cannabis to enhance programming, but did not observe any effect. 

Finally, while we observed little quantitative evidence of negative experiences of cannabis use while programming, we did observe a few qualitative reports. For example, one participant stated: ``I generally don't use [cannabis] while programming because [\dots] it affects my short-term memory, which is a huge part of programming for me [\dots] My managers wouldn't have an issue with me using cannabis during my job, but I do have an issue just due to the nature of cannabis.'' Similarly, another participant stated ``I wanted to see if [cannabis] would help, but all it did was make it harder to keep track of what I was doing. I'm glad I tried it, but I wouldn't do it again.'' These quotes show that even among cannabis-using programmers, there is a wide array of experiences and reactions, a variance that invites further and more in-depth qualitative analysis.

\begin{table}[t]
\small

\begin{tabular}{@{}lrrr@{}}
\toprule
 \textbf{Reason for cannabis use while programming} & \textbf{Cat} & \textbf{Count} & \textbf{\%} \\ \midrule
\rowcolor[HTML]{BCEDA7} 
To make programming tasks more enjoyable         & E   & 148 & 60.9 \\
\rowcolor[HTML]{DAE8FC} 
To think of more creative programming solutions  & P   & 128 & 52.7 \\
\rowcolor[HTML]{DAE8FC} 
To get in a programming zone                     & P   & 117 & 48.2 \\
\rowcolor[HTML]{BCEDA7} 
To make programming-related tasks less tedious   & E   & 103 & 42.4 \\
\rowcolor[HTML]{DAE8FC} 
To enhance brainstorming                         & P   & 96  & 39.5 \\
\rowcolor[HTML]{DAE8FC} 
To focus on programming-related tasks            & P   & 80  & 32.9 \\
\rowcolor[HTML]{DAE8FC} 
To gain insight or understanding                 & P   & 79  & 32.5 \\
\rowcolor[HTML]{FFFFC7} 
To help with work-related anxiety                & W   & 67  & 27.6 \\
\rowcolor[HTML]{BCEDA7} 
To have fun in social programming settings       & E   & 35  & 14.4 \\
Other (please describe)                          & N/A & 33  & 13.6 \\
\rowcolor[HTML]{FFFFC7} 
To mitigate programming-related pain             & W   & 32  & 13.2 \\
\rowcolor[HTML]{FFFFC7} 
To help with work-related social anxiety         & W   & 32  & 13.2 \\
\rowcolor[HTML]{BCEDA7} 
To improve social interactions in the workplace  & E   & 18  & 7.4  \\
\rowcolor[HTML]{FFFFC7} 
For non-programming related medical conditions   & W   & 17  & 7.0  \\
\rowcolor[HTML]{FFFFC7} 
To help with work-related migraines              & W   & 12  & 4.9  \\
\rowcolor[HTML]{FFFFC7} 
I am unable to think as clearly without cannabis & W   & 11  & 4.5  \\ \bottomrule
\end{tabular}
    \caption{``Why do you use cannabis while programming, coding, or completing other software engineering-related tasks?'' Participants could select multiple choices. ``Cat'' delineates a particular motivation as programming enhancement (P), enjoyment (E), or wellness (W). \% is out of the 243 who selected at least one option. }    \label{tab:whyUse} \end{table}


\textbf{Implications}---As a result of these conflicting experiences, our findings raise the question of whether the programming-related benefits of cannabis perceived by some developers actually translate to verifiable programming enhancement. For example, even though many participants report using cannabis to enhance programming creativity, it is unclear if this actually occurs. We note this concern was raised by some non-cannabis using programmers. For instance, one such programmer wrote that they ``had a series of developers work for [them] that used cannabis to varying degrees. All of them fully believed that it made them better engineers, that is sparked creativity and capability. From the outside however the results have consistently not been that way. Not just from less code productivity, but far more often inability to work as well with others, and code quality issues.'' Thus, our results motivate future observational study of the effects of cannabis on programming. This motivation and considerations surrounding any such study are discussed further in Section~\ref{sec:discussion}. 

\begin{framed}{
    \noindent We found that programmers who use cannabis were more likely to be motivated by potential programming ability enhancement or programming enjoyment than wellness-related reasons. 
    This pattern was observed in both open source developers and university participants, and motivates future work investigating if the perceived programming benefits of cannabis manifest in a more rigorous observational study. 
}
\end{framed}

\subsection{RQ4: Perception of Cannabis Use}

We also investigate how programmers perceive cannabis use. Understanding this perception is important because if perception varies from actual usage, this may result in sub-optimal cannabis-related policies or biases in programming environments. We investigate cannabis perceptions in general and programming-specific contexts. For the former, we compare to national cannabis attitudes surveys. For the latter, we analyze attitude differences between programming students, employees, and software managers. We then compare any differences to each group's cannabis usage.  

\textbf{General Cannabis Perceptions---}Programmers in our sample have more positive attitudes towards cannabis than the population overall. For example, 91\% of our participants say that marijuana use should be legal for both recreational and medicinal use compared to 60\% of the general United States population in 2021~\cite{green2021pew}. Similarly, only 5\% of our population views smoking marijuana once or twice a week to be of ``great risk'' as opposed to 29\% of the US population~\cite{NSDUH}. This difference is likely explained by the demographic differences in age, gender, and political leaning between programmers in our sample and the population overall (see Section~\ref{sec:pop_context}).

\textbf{Programming and Cannabis Perceptions Setup---}To understand cannabis perceptions in a programming context, we first ask participants to rate their approval or disapproval of someone who uses cannabis while working with them on software engineering tasks such as programming, brainstorming, debugging, or security testing, on a 5-point Likert scale (from -2 for disapproval to +2 for approval). We average these responses into an overall ``cannabis approval score''. This approval/disapproval format was adapted from previous research on cannabis attitudes~\cite{bachman1988explaining}.
Second, we indirectly asked about cannabis use \textit{visibility} by asking if respondents knew of a colleague who regularly used cannabis while programming on a 5-point Likert scale (from -2 for a solid no to +2 for a solid yes).

The exact wording varied for student, employee, or manager participants. Differentiating between groups admits investigating more nuanced perception differences. For example, we ask employees to indicate if they think their manager would disapprove of cannabis use while we ask managers if they actually would disapprove. Even though this is not a perfect comparison, as managers in our sample do not necessarily manage employees in our sample, it still allows a first investigation into perception differences between these two groups overall. Table~\ref{tab:perception_questions} lists our population-specific
questions.

\begin{table*}[t]
\small
\begin{tabular}{@{}llrrrr@{}}
\toprule

Q ID &
  Perception Question &
  Answer Choices &
  \begin{tabular}[c]{@{}r@{}}Population\end{tabular} &
  \begin{tabular}[c]{@{}r@{}}Responses\end{tabular} & Score\\ \midrule
  &\textit{Perceived approval/disapproval of cannabis use while programming:}\\ 

  \midrule
\rowcolor[HTML]{EFEFEF} 
1 &
  \begin{tabular}[c]{@{}l@{}}Would you approve/disapprove of a student partner \\ 
  using cannabis while \rule{0.75cm}{0.15mm} for school-related tasks?\end{tabular} &
  \begin{tabular}[c]{@{}r@{}}5-point likert: \\ disapprove/approve\end{tabular} &
  \begin{tabular}[c]{@{}r@{}}Programming\\Students\end{tabular} &
  237 & -0.25\\ 
2 &
  \begin{tabular}[c]{@{}l@{}}Would you approve/disapprove of a colleague\\ 
  using cannabis while \rule{0.75cm}{0.15mm} for work-related tasks?\end{tabular} &
  \begin{tabular}[c]{@{}r@{}}5-point likert: \\ disapprove/approve\end{tabular} &
  \begin{tabular}[c]{@{}r@{}}Non-manager\\ Employees\end{tabular} &
  329 & -0.36 \\
\rowcolor[HTML]{EFEFEF} 
3 &
  \begin{tabular}[c]{@{}l@{}}Would your manager approve/disapprove of you \\ using cannabis 
  while \rule{0.75cm}{0.15mm} for work-related tasks?\end{tabular} &
  \begin{tabular}[c]{@{}r@{}}5-point likert: \\ disapprove/approve\end{tabular} &
  \begin{tabular}[c]{@{}r@{}}Non-manager\\ Employees\end{tabular} &
  306 & -1.04\\
4 &
  \begin{tabular}[c]{@{}l@{}}Would you approve/disapprove of an employee you supervise\\ 
  using cannabis while \rule{0.75cm}{0.15mm} for work-related tasks?\end{tabular} &
  \begin{tabular}[c]{@{}r@{}}5-point likert: \\ disapprove/approve\end{tabular} &
  \begin{tabular}[c]{@{}r@{}}Programming\\ Managers\end{tabular} &
 189 & -0.51
 \\ \midrule
 & \textit{Perceived visibility of cannabis use while programming:} &
   &
   \\ \midrule
\rowcolor[HTML]{EFEFEF} 
5 &
  \begin{tabular}[c]{@{}l@{}}Do you know a fellow student who regularly uses cannabis \\ 
  while completing programming-related tasks?\end{tabular} &
  \begin{tabular}[c]{@{}r@{}}5-point likert \\ (no...unsure...yes)\end{tabular} &
  \begin{tabular}[c]{@{}r@{}}Programming \\ Students\end{tabular} &
  236 & 0.36\\
6 &
  \begin{tabular}[c]{@{}l@{}}Do you know a colleague at your workplace who regularly \\ uses cannabis while completing programming-related?\end{tabular} &
  \begin{tabular}[c]{@{}r@{}}5-point likert \\ (no...unsure...yes)\end{tabular} &
   \begin{tabular}[c]{@{}r@{}}Programming \\ Workers\end{tabular}&
  519 & -0.25\\
\rowcolor[HTML]{EFEFEF} 
7 &
  \begin{tabular}[c]{@{}l@{}}Do you know a manager at your workplace who regularly \\ uses cannabis while completing programming-related tasks?\end{tabular} &
  \begin{tabular}[c]{@{}r@{}}5-point likert \\ (no...unsure...yes)\end{tabular} &
  \begin{tabular}[c]{@{}r@{}}Programming \\ Workers\end{tabular} &
  519 & -0.72\\ \bottomrule
\end{tabular}
\caption{Student, employee, and manager perception questions (wordings modified slightly for clarity). For Q ID 1--4, participants gave approval for programming, brainstorming, debugging, and security testing, then aggregated for an overall participant score. The score column represents the average Likert score (from -2 to +2) for all ``Response'' participants.  
}
\label{tab:perception_questions}
\end{table*}

\textbf{Perceptions of Professionals vs. Students---}For student programmers, we considered those who were currently a student in computer science, software engineering, or another programming-related field. For professional programmers, we consider those non-students who were either full-time employees, part-time employees, or self-employed with a programming job.
We hypothesized that students would approve more of cannabis use. However, we find no evidence of a difference between employee approval of colleagues (question 2) and student approval of fellow students (question 1) for using cannabis while completing programming tasks: the average from both groups was between neutral (0) and slight disapproval (-1) (-0.25 for student on students and -0.36 for employee on colleagues, $p=0.26$). This indicates that perceptions of cannabis use while programming are similar in academic and professional contexts.

For visibility, however, we do observe a significant difference. Responses to questions 5 and 6 show that students are significantly more likely to know another student who regularly uses cannabis while programming than a professional programmer is to know a colleague who does the same ($p<0.0001$): 48\% of students report knowing or probably knowing a fellow student who regularly uses cannabis while programming compared to only 23\% of professional programmers.
However, we observe no significant differences in cannabis usage prevalence or frequency between the two groups, though fewer students than professionals report cannabis use while programming (32\% vs 38\%, $p=0.09$). This finding may represent cultural differences despite similar approval and usage levels between the two groups, differences perhaps driven by higher levels of support for cannabis legalization among younger Americans~\cite{green2021pew} or the fact that some students were under the age of 21, the most common age limit for legal cannabis use in the United States.

\textbf{Perceptions of Managers vs. Employees---}For programming employees, we consider full-time, part-time, and self-employees with a programming job. Managers were those who reported they were managers. 
The average level of \emph{expected} managers disapproval of cannabis use by employees was between slight and strong. However, managers reported an \emph{actual} disapproval level between neutral and slight---a significant difference (questions 3 and 4, $p<0.0001$). We found no significant difference between manager disapproval of their supervisees and employee disapproval of their colleagues (questions 2 and 4, $p=0.15$), both reporting between neutral and slight disapproval. We also found no significant cannabis usage differences between employees and managers. Additionally, managers rarely report witnessing negative cannabis-related effects: while 27\% (51/189) of managers suspect a supervisee uses cannabis, less than 3\% (5/189) report that such programmers were less productive, and only one reported reprimanding an employee for cannabis use. Thus, our results indicate a perception mismatch of programming cannabis use between employees and managers: employees expect managers to disapprove of such use more than they actually do. 

This mismatch between managers and employees is further compounded by a difference in cannabis-use visibility. Professional programmers are more likely to know or probably know a colleague who regularly uses cannabis while programming than they are to know a manager who regularly uses cannabis while programming (23\% vs. 8\%, $p<0.0001$). One potential implication is that if managers are ambivalent on cannabis use while programming, then corporate policies might be adjusted to avoid repercussions for cannabis use as long as work performance remains reasonable. 

\begin{framed}{
    \noindent We found that US-based programmers have more favorable views on general cannabis use and legalization than the American population overall. In programming-specific contexts, we found differences in perception and/or visibility of cannabis use while programming between programming students, employees, and managers despite finding no significant differences in cannabis usage between the three groups. For example, we found that managers disapprove of cannabis less than employees expect them to. These results may indicate a mismatch between perception and reality.
}
\end{framed}

\section{Discussion}
\label{sec:discussion}

 
In this paper, we report that cannabis frequently interacts with programming in both educational and professional software engineering contexts. However, our work is also only a first step toward understanding this intersection. In this discussion, we consider the impacts of our findings on programming workplace culture and policies. We also discuss questions raised by our results and potential directions for future work including focus groups and observational studies. We conclude with a broader discussion of the intersection of cannabis and software engineering in general.


 \textbf{Cannabis Use and Company Policies---}Our findings have implications for software company anti-drug policies. While perhaps less prevalent than in other industries (notably, many FANG (Facebook, Amazon, Netflix, Google) companies do not drug-test employees), anti-cannabis regulations remain common in programming environments (e.g.,~\cite[p.~12]{ciscoHandbook}
 and~\cite[pp.~12--13]{ibmHandbook}), policies often enforced through drug testing: 29\% of our participants reported taking a drug test for a programming job. 
 
 At the same time, we found cannabis use is common among programmers, with 35\% of our sample having used cannabis while programming, 34\% of which sometimes do so for work-related tasks (especially during remote work). Cannabis-using programmers typically use THC products, the chemical detected by most drug tests. Furthermore, perceptions of cannabis use while programming are mostly ambivalent from both managers and employees, and we observe limited reports of cannabis use negatively impacting programming work-places: while 27\% of programming managers suspect cannabis use by a supervisee, less than 3\% report that those programmers were less productive, and only one manager in our sample reported reprimanding an employee for cannabis use. 
 
 Thus, our results indicate that software company anti-drug policies may conflict with the preferred practice of many developers, a dissonance that can lead to 
 smaller application pools and hiring difficulties for drug-testing jobs (as evinced by the 2014 FBI cybersecurity hiring shortage~\cite{bbcFBI2014}). Our results support additional study and reassessment of extant anti-drug policies for programming jobs including further contextualization on when and how often programmers currently take drug tests for programming-related jobs\footnote{Unfortunately, we did not collect data regarding the types of drug testing experienced by programmers (e.g., once during hiring vs. continuous spot-checking), data that would provide useful context. We encourage future work to explore this gap more directly.} and if such policies are actually necessary or effective. 


\textbf{Outstanding Questions and Future Directions---}One direct question is
functional: does cannabis use while programming \emph{actually affect code quality}, and if so, how? From creativity to readability to defect density, there are a number of purported effects of cannabis use on code. This is particularly compelling because programming cannabis use is often motivated by the desire to enhance programming skills such as brainstorming and focus (see Section~\ref{sec:rq_motivation}).
Observational studies, in which programmers complete various software tasks with and without cannabis, provide one way to investigate such questions. Both functional
outcomes and also medical markers (e.g., eye tracking, blood testing, etc.) are relevant. 
Such studies are currently legally challenging in most parts of world.
However, cannabis laws are increasing in leniency, and in the US, both CBD and THC have synthetic versions approved for medicinal use by federal regulatory boards. Finally, some observational studies have been conducted with real-world cannabis. For example, Bidwell \emph{et al.} measured impairment with state-legalized cannabis (comparing THC dosages) using a series of cognitive tests~\cite{bidwell2020association}, encouraging the pursuit of similar observational studies in a software context.

Other research techniques may be more immediately applicable. Qualitative studies employing interviews or focus groups could deepen understanding of cannabis programming culture. Such studies could also help identify if and when anti-drug polices impact job application or hiring decisions. Other directions include additional survey-based research on other programmers sub-populations (such as workers in public vs. private companies) or on programmers' cannabis usage in countries beyond the United States.

\textbf{Broader Software Engineering Impacts---}Our work also has broader implications on software engineering ecosystems and cannabis-connected human-computer interactions. The  intersection of cannabis use and human-computer interaction remains unexplored, in both general and software contexts. As cannabis becomes more acceptable, it may interact more with socio-technical issues including accessibility, end-user security, data privacy, and programming identity. This possibility was emphasized with ``CHI-nnabis'', a recent panel at \textit{Computer Human Interaction} (CHI) calling for more research at the intersection of cannabis and technology~\cite{chinnabis}.


While we focused on cannabis use, there may be other understudied mind-altering substances that interact with programming. For instance, while psychedelics had connections to the nascent commercial computing industry~\cite{walsh2011drugs, markoff2005dormouse}, little is known about current prevalence in software engineering, let alone the accuracy of any perceived benefits. Similarly, while college students often abuse prescription stimulants as ``study drugs''~\cite{varga2012adderall}, little is known about the use of and culture surrounding such drugs by computing students or in a programming context. Overall, we hope our work spurs research on the intersection of programming and mind-altering substances from multiple angles, such as company drug policies, programming productivity, and socio-technical considerations. 

\section{Limitations and Threats to Validity}
\label{sec:limitations}


Other limitations include self-selection bias and data quality concerns, common for survey research in general and for self-reported drug research in particular~\cite{harrison1997validity}. Due to interest, cannabis-using programmers may be more likely than non-users to participate. Simultaneously, cannabis users may be less willing to report use due to retaliation worries. To mitigate these biases, we made it clear that prior cannabis use was not necessary (see Section~\ref{sec:design}), and our survey was anonymous and confidential to encourage honesty. To check for self-selection bias, we validated that key demographics in our sample match those of our recruitment populations (see Section~\ref{sec:pop_context}). We also note that online cannabis use surveys typically produce high quality and internally consistent data~\cite{ramo2012reliability}. Even so, these biases are important when interpreting our findings and may result in our findings overestimating the prevalence of cannabis use while programming.

One additional limitation is timing during COVID-19: our findings may be influenced by this period of primarily remote work. We mitigate this threat by asking about cannabis-use behavior changes resulting from the onset of COVID-19 (see Section~\ref{sec:prog_context}). We note, however, that while this is a potential limitation, it also admits timely and useful analysis in light of our finding that remote work increases cannabis use while programming for work-related tasks.


\section{Conclusion}
In this paper, we presented results of \textbf{the first empirical study of cannabis's prevalence, perceptions, and usage motivations in programming environments}. To do so, we conducted a survey of~\popSize{} programmers (including 450 full-time professional developers) recruited from open source, university, and social media programming communities. We found that some programmers regularly  use cannabis while programming (18\% of our sample do so at least once a month), many choosing to use cannabis for both personal and work-related projects. Furthermore, we find that cannabis use while programming is primarily motivated by perceived enhancement to programming-related skills and increased enjoyment rather than by medicinal reasons. Finally, we find that programming employees, managers, and students use cannabis while programming at similar rates, despite differences in cannabis perceptions and visibility.

Such cannabis usage, however, is in conflict with anti-drug policies currently enacted for many software engineering jobs: 29\% of our sample reported they had taken a drug test for a programming-related job, a hiring practice that may limit developer application pools. Thus, our results have implications for programming workplaces that currently have anti-drug policies and motivate future research into the effects of cannabis use while programming.


\section*{Acknowledgements}
We acknowledge the partial support of the NSF (CCF 1908633, CCF 1763674) and a 
Google Faculty Research Award. Dr. Boehnke’s effort on this publication was partially supported by the National Institute on Drug Abuse of the National Institutes of Health under Award Number K01DA049219 (KFB). The content is solely the responsibility of the authors and does not necessarily represent the official views of the National Institutes of Health.
Additionally, we extend our thanks to Li Morrow, our point of contact at the University of Michigan's IRB, for her help with answering our questions regarding ethical considerations for the design of this study. 

\bibliographystyle{ACM-Reference-Format}
\balance 
\bibliography{endres_bib}

\end{document}